\documentclass[12pt]{article}
 \usepackage{epsfig}
 \def\be{\begin{equation}}
 \def\ee{\end{equation}}
 \def\bea{\begin{eqnarray}}
 \def\eea{\end{eqnarray}}
 \usepackage{graphicx}

 \catcode`\@=11
 \def\lsim{\mathrel{\mathpalette\@versim<}}
 \def\gsim{\mathrel{\mathpalette\@versim>}}
 \def\@versim#1#2{\vcenter{\offinterlineskip
 \ialign{$\m@th#1\hfil##\hfil$\crcr#2\crcr\sim\crcr } }}
 \catcode`\@=12

 \catcode`@=12
\begin{document}
 \thispagestyle{empty}
 \begin{flushright}
 UCRHEP-T565\\
 April 2016\
 \end{flushright}
 \vspace{0.6in}
 \begin{center}
 {\LARGE \bf Gauge $U(1)$ Dark Symmetry and\\
 Radiative Light Fermion Masses\\}
 \vspace{1.0in}
 {\bf Corey Kownacki$^1$ and Ernest Ma$^{1,2,3}$\\}
 \vspace{0.2in}
 {\sl $^1$ Department of Physics and Astronomy,\\ 
 University of California, Riverside, California 92521, USA\\}
 \vspace{0.1in}
 {\sl $^2$ Graduate Division,  
 University of California, Riverside, California 92521, USA\\}
 \vspace{0.1in}
 {\sl $^3$ HKUST Jockey Club Institute for Advanced Study,\\ 
 Hong Kong University of Science and Technology, Hong Kong, China\\}

 \end{center}
 \vspace{1.0in}

\begin{abstract}\
A gauge $U(1)$ family symmetry is proposed, spanning the quarks and leptons 
as well as particles of the dark sector.  The breaking of $U(1)$ to $Z_2$ 
divides the two sectors and generates one-loop radiative masses for the 
first two families of quarks and leptons, as well as all three neutrinos.  
We study the phenomenological implications of this new connection between 
family symmetry and dark matter.  In particular, a scalar or pseudoscalar 
particle associated with this $U(1)$ breaking may be identified with the 
750 GeV diphoton resonance recently observed at the Large Hadron Collider 
(LHC).
\end{abstract}

\newpage
 \baselineskip 24pt

\noindent \underline{\it Introduction}~:\\
In any extension of the standard model (SM) of particle interactions to 
include dark matter, a symmetry is usually assumed, which distinguishes 
quarks and leptons from dark matter.  For example, the simplest choice is 
$Z_2$ under which particles of the dark sector are odd and those of the 
visible sector are even.  Suppose $Z_2$ is promoted to a gauge $U(1)$ 
symmetry, then the usual assumption is that it will not affect ordinary 
matter.  These models all have a dark vector boson which 
couples only to particles of the dark sector.
\begin{table}[htb]
\caption{Particle content of proposed model of gauge $U(1)$ dark symmetry.}
\begin{center}
\begin{tabular}{|c|c|c|c|c|c|}
\hline
particles & $SU(3)_C$ & $SU(2)_L$ & $U(1)_Y$ & $U(1)_D$ & $Z_2$ \\
\hline
$Q = (u,d)$ & 3 & 2 & 1/6 & $0,0,0$ & + \\
$u^c$ & $3^*$ & 1 & $-2/3$ & $1,-1,0$ & + \\
$d^c$ & $3^*$ & 1 & 1/3 & $-1,1,0$ & + \\
\hline
$L = (\nu,e)$ & 1 & 2 & $-1/2$ & $0,0,0$ & + \\
$e^c$ & 1 & 1 & 1 & $-1,1,0$ & + \\
\hline
$\Phi = (\phi^+,\phi^0)$ & 1 & 2 & $1/2$ & $0$ & + \\
$\sigma_1$ & 1 & 1 & 0 & $1$ & + \\
$\sigma_2$ & 1 & 1 & 0 & $2$ & + \\
\hline
$N,N^c$ & 1 & 1 & 0 & $1/2,-1/2$ & $-$ \\  
$S,S^c$ & 1 & 1 & 0 & $-3/2,3/2$ & $-$ \\  
\hline
$(\eta^0,\eta^-)$ & 1 & 2 & $-1/2$ & $1/2$ & $-$ \\ 
$\chi^0$ & 1 & 1 & $0$ & $1/2$ & $-$ \\  
$\chi^-$ & 1 & 1 & $-1$ & $-1/2$ & $-$ \\  
\hline
$(\xi^{2/3},\xi^{-1/3})$ & 3 & 2 & 1/6 & $1/2$ & $-$ \\ 
$\zeta^{2/3}$ & 3 & 1 & 2/3 & $-1/2$ & $-$ \\ 
$\zeta^{-1/3}$ & 3 & 1 & $-1/3$ & $-1/2$ & $-$ \\ 
\hline
\end{tabular}
\end{center}
\end{table}

In this paper, it is proposed instead that a gauge $U(1)$ extension of the 
SM spans both ordinary and dark matter.  It is in fact also a horizontal 
family symmetry.  It has a number of interesting 
consequences, including the radiative mass generation of the first two 
families of quarks and leptons, and a natural explanation of the 750 GeV 
diphoton resonance recently observed~\cite{atlas15,cms15} at the Large 
Hadron Collider (LHC). 

\noindent \underline{\it New Gauge $U(1)_D$ Symmetry}~:\\
The framework that radiative fermion masses and dark matter are related 
has been considered previously~\cite{m14}.  Here it is further proposed 
that families are distinguished by the connecting dark symmetry.  
In Table 1 we show how they transform under $U(1)_D$ as well as the 
other particles of the dark sector.  The $U(1)_D$ symmetry is broken 
spontaneously by the vacuum expectation value $\langle \sigma_{1,2} \rangle 
= u_{1,2}$ to an exactly conserved $Z_2$ which divides the two sectors.
\begin{figure}[htb]
\vspace*{-3cm}
\hspace*{-3cm}
\includegraphics[scale=1.0]{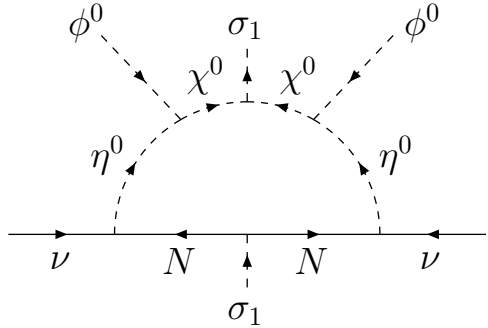}
\vspace*{-20.5cm}
\caption{One-loop neutrino mass from trilinear couplings.}
\end{figure}
\begin{figure}[htb]
\vspace*{-3cm}
\hspace*{-3cm}
\includegraphics[scale=1.0]{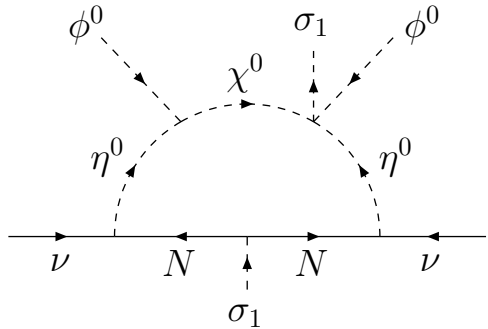}
\vspace*{-20.5cm}
\caption{One-loop neutrino mass from trilinear and quadrilinear couplings.}
\end{figure}

The gauge $U(1)_D$ symmetry is almost absent of axial-vector anomalies for 
each 
family.  The $[SU(3)]^2 U(1)_D$ anomaly is zero from the cancellation between 
$u^c$ and $d^c$.  The $[SU(2)]^2 U(1)_D$ anomaly is zero because $Q$ and $L$ 
do not transform under $U(1)_D$.  The $[U(1)_Y]^2 U(1)_D$ and 
$U(1)_Y [U(1)_D]^2$ anomalies are cancelled among $u^c$, $d^c$, and $e^c$, i.e.
\begin{eqnarray}
&& 3 \left( -{2 \over 3} \right)^2 (1) + 3 \left( {1 \over 3} \right)^2 (-1) 
+ (1)^2 (-1) = 0, \\
&& 3 \left( -{2 \over 3} \right)(1)^2 + 3 \left( {1 \over 3} \right) (-1)^2 + 
(1)(-1)^2 = 0.
\end{eqnarray}
The $[U(1)_D]^3$ anomaly is not zero for either the first or second family, 
but is cancelled between the two.  This is thus a generalization of the 
well-known anomaly-free $L_e - L_\mu$ gauge symmetry~\cite{hjlv91} to the 
difference of $B-L-2Y$ between the first two families.

\begin{figure}[htb]
\vspace*{-3cm}
\hspace*{-3cm}
\includegraphics[scale=1.0]{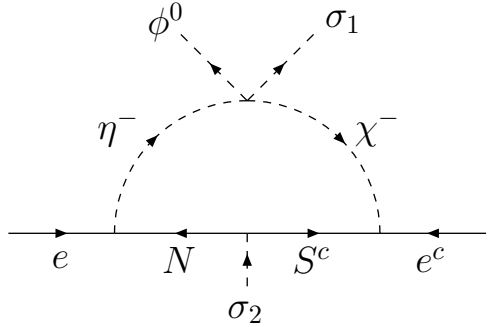}
\vspace*{-20.5cm}
\caption{One-loop electron mass.}
\end{figure}
\begin{figure}[htb]
\vspace*{-3cm}
\hspace*{-3cm}
\includegraphics[scale=1.0]{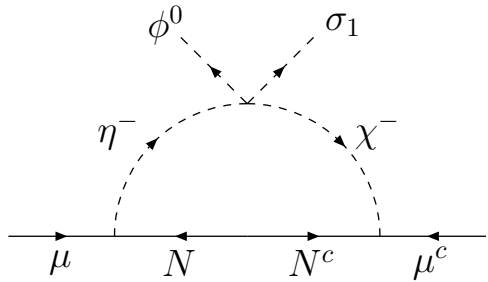}
\vspace*{-21.5cm}
\caption{One-loop muon mass.}
\end{figure}

\newpage

\noindent \underline{\it Radiative Masses for Neutrinos and the First 
and Second Families}~:\\
At tree level, only $t,b,\tau$ acquire masses from $\langle \phi^0 \rangle 
= v$ as in the SM.  The first two families are massless because of the 
$U(1)_D$ symmetry.  Neutrinos acquire one-loop masses through the scotogenic 
mechanism~\cite{m06} as shown in Figs.~1 and 2.
With one copy of $(N,N^c)$, only one neutrino becomes massive.  To have three 
massive scotogenic neutrinos, three copies of $(N,N^c)$ are needed.
The one-loop electron and muon masses are shown in Figs.~3 and 4.
Note that at least two copies of $(N,N^c)$ are needed for two charged-lepton 
masses.
The mass matrix spanning $(N,N^c,S,S^c)$ is of the form
\begin{equation}
{\cal M}_{N,S} = \pmatrix{f_1 u_1 & m_N & f_3 u_1 & f_5 u_2 \cr 
m_N & f_2 u_1 & f_6 u_2 & f_4 u_1 \cr f_3 u_1 & f_6 u_2 & 0 & m_S \cr 
f_5 u_2 & f_4 u_1 & m_S & 0}.
\end{equation}
Note that the $f_{1,2,3,4} u_1$ terms break lepton number by two units, 
whereas the $f_{5,6} u_2$ terms do not.
\begin{figure}[htb]
\vspace*{-3cm}
\hspace*{-3cm}
\includegraphics[scale=1.0]{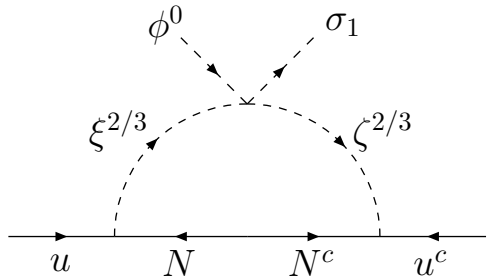}
\vspace*{-21.5cm}
\caption{One-loop $u$ quark mass.}
\end{figure}
\begin{figure}[htb]
\vspace*{-3cm}
\hspace*{-3cm}
\includegraphics[scale=1.0]{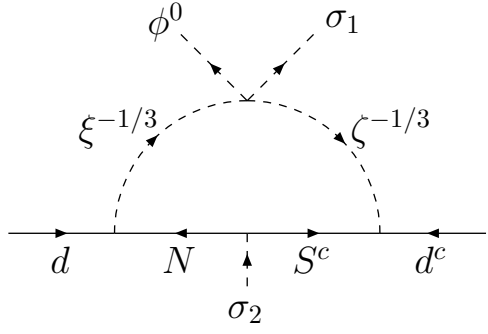}
\vspace*{-20.5cm}
\caption{One-loop $d$ quark mass.}
\end{figure}
Lepton number $L=1$ may be assigned to $e,\mu,\tau,N,S$ and $L=-1$ to 
$e^c,\mu^c,\tau^c,N^c,S^c$.  It is broken down to lepton parity $(-1)^L$ 
only by neutrino masses.
The analogous one-loop $u$ and $d$ quark masses are shown in Figs.~5 and 6.
Because the second family has opposite $U(1)_D$ charge assignments relative 
to the first, the $c$ and $s$ quarks reverse the roles of $u$ and $d$.
Two copies of $(S,S^c)$ are needed to obtain the most general quark mass 
matrices for both the $u$ and $d$ sectors.

To evaluate the one-loop diagrams of Figs.~1 to 6, we note first that each 
is a sum of simple diagrams with one internal fermion line and one internal 
scalar line.  Each contribution is infinite, but the sum is finite.  There 
are 10 neutral Majorana fermion fields, spanning 3 copies of $N,N^c$ and 
2 copies of $S,S^c$.  We denote their mass eigenstates as $\psi_k$ with 
mass $M_k$.  There are 4 real scalar fields, spanning $\sqrt{2} Re(\eta^0)$, 
$\sqrt{2} Im(\eta^0)$, $\sqrt{2} Re(\chi^0)$, $\sqrt{2} Im(\chi^0)$.  We 
denote their mass eigenstates as $\rho^0_l$ with mass $m_l$.  
In Figs.~1 and 2, let the 
$\nu_{i} \psi_k \bar{\eta}^0$ coupling be $h^\nu_{ik}$, then the radiative 
neutrino mass matrix is given by~\cite{m06}
\begin{equation}
({\cal M}_\nu)_{ij} = \sum_k {h^\nu_{ik} h^\nu_{jk} M_k \over 16 \pi^2} 
\sum_l [(y^R_l)^2 F(x_{lk}) - (y^I_l)^2 F(x_{lk})],
\end{equation}
where $\sqrt{2} Re(\eta^0) = \sum_l y^R_l \rho^0_l$, $\sqrt{2} Im(\eta^0) = 
\sum_l y^I_l \rho^0_l$, with $\sum_l (y^R_l)^2 = \sum_l (y^I_l)^2 = 1$, 
$x_{lk} = m_l^2/M_k^2$, and the function $F$ is given by
\begin{equation}
F(x) = {x \ln x \over x-1}.
\end{equation}
There are two charged scalar fields, spanning $\eta^\pm,\chi^\pm$.  We 
denote their mass eigenstates as $\rho^+_r$ with mass $m_r$. 
In Fig.~3, let the $e_L \psi_k \eta^+$ and the $e^c_L \psi_k \chi^-$ 
couplings be $h^e_k$ and $h^{e^c}_k$, then
\begin{equation}
m_e = \sum_k {h^e_k h^{e^c}_k M_k \over 16 \pi^2} \sum_r y^\eta_r y^\chi_r 
F(x_{rk}),
\end{equation}
where $\eta^+ = \sum_r y^\eta_r \rho^+_r$, $\chi^+ = \sum_r y^\chi_r 
\rho^+_r$, with $\sum_r (y^\eta_r)^2 = \sum_r (y^\chi_r)^2 = 1$ and 
$\sum_r y^\eta_r y^\chi_r = 0$.  A similar expression is obtained for 
$m_\mu$, as well as the light quark masses.

\noindent \underline{\it Tree-Level Flavor-Changing Neutral Couplings}~:\\
Since different $U(1)_D$ charges are assigned to $(u^c,c^c,t^c)$ as well 
as $(d^c,s^c,b^c)$, there are unavoidable flavor-changing neutral currents. 
They can be minimized by the following assumptions.  Let the two $3 \times 3$ 
quark mass matrices linking $(u,c,t)$ to $(u^c,c^c,t^c)$ and $(d,s,b)$ to 
$(d^c,s^c,b^c)$ be of the form
\begin{equation}
{\cal M}_u = U_L^{(u)} \pmatrix{m_u & 0 & 0 \cr 0 & m_c & 0 \cr 0 & 0 & m_t}, 
~~~ {\cal M}_d = U_L^{(d)} \pmatrix{m_d & 0 & 0 \cr 0 & m_s & 0 \cr 
0 & 0 & m_b}, 
\end{equation}
where $U_{CKM} = (U_L^{(u)})^\dagger U_L^{(d)}$ is the quark charged-current 
mixing matrix.  However, since $Z_D$ does not couple to left-handed quarks, 
and its couplings to right-handed quarks have been chosen to be diagonal 
in their mass eigenstates, flavor-changing neutral currents are absent 
in this sector.  Of course, they will appear in the scalar sector, and 
further phenomenological constraints on its parameters will apply.

\noindent \underline{\it $Z_D$ Gauge Boson}~:\\
As $\sigma_{1,2}$ acquire vacuum expectation values $u_{1,2}$ respectively, 
the $Z_D$ gauge boson obtains a mass given by
\begin{equation}
m^2_{Z_D} = 2 g_D^2 (u_1^2 + 4 u_2^2).
\end{equation}
Since $\sigma_{1,2}$ do not transform under the SM, and $\Phi$ does not 
under $U(1)_D$, there is no mixing between $Z_D$ and $Z$.  Using Table 1 
and assuming that all new particles are lighter than $Z_D$, the branching 
fraction of $Z_D$ to $e^-e^+ + \mu^- \mu^+$ is estimated to be 0.07. 
The $c_{u,d}$ coefficients used in the experimental 
search~\cite{atlas14,cms14} of $Z_D$ are then
\begin{equation}
c_u = c_d = g_D^2~(0.07).
\end{equation}
For $g_D=0.3$, a lower bound of about 3.1 TeV on $m_{Z_D}$ is obtained 
from LHC data based on the 7 and 8 GeV runs.  For our subsequent discussion, 
let $u_1 = 1$ TeV, $u_2 = 4$ TeV, then $m_{Z_D} = 3.4$ TeV.  Note that 
$Z_D$ does not couple to the third family, so if $\bar{t} t$, $\bar{b} b$, 
or $\tau^+ \tau^-$ final states are observed, this model is ruled out.

\noindent \underline{\it Scalar Sector}~:\\
There are three scalars with integral charges under $U(1)_D$, i.e. $\Phi$ and 
$\sigma_{1,2}$.  Whereas $\langle \phi^0 \rangle = v$ breaks the electroweak 
$SU(2)_L \times U(1)_Y$ gauge symmetry as in the SM, $\langle \sigma_{1,2} 
\rangle = u_{1,2}$ break $U(1)_D$ to $Z_2$, with all those particles with 
half-integral $U(1)_D$ charges becoming odd under this exactly conserved 
dark $Z_2$ parity.  The relevant scalar potential is given by
\begin{eqnarray}
V &=& \mu_0^2 \Phi^\dagger \Phi + m_1^2 \sigma_1^* \sigma_1 + m_2^2 \sigma_2^* 
\sigma_2 + m_{12} \sigma_1^2 \sigma_2^* + m_{12} (\sigma_1^*)^2 \sigma_2 
\nonumber \\ 
&+& {1 \over 2} \lambda_0 (\Phi^\dagger \Phi)^2 + {1 \over 2} \lambda_1 
(\sigma_1^* \sigma_1)^2 + {1 \over 2} \lambda_2 (\sigma_2^* \sigma_2)^2 
+ \lambda_3 (\sigma_1^* \sigma_1)(\sigma_2^* \sigma_2) \nonumber \\ 
&+& \lambda_4 
(\Phi^\dagger \Phi)(\sigma_1^* \sigma_1) + \lambda_5 (\Phi^\dagger \Phi) 
(\sigma_2^* \sigma_2),
\end{eqnarray}
where $m_{12}$ has been rendered real by absorbing the relative phase between 
$\sigma_{1,2}$.  The conditions for $v$ and $u_{1,2}$ are
\begin{eqnarray}
0 &=& \mu_0^2 + \lambda_0 v^2 + \lambda_4 u_1^2 + \lambda_5 u_2^2, \\ 
0 &=& m_1^2 + \lambda_1 u_1^2 + \lambda_3 u_2^2 + \lambda_4 v^2 
+ 2 m_{12} u_2, \\ 
0 &=& m_2^2 + \lambda_2 u_2^2 + \lambda_3 u_1^2 + \lambda_5 v^2 
+ m_{12} u_1^2/u_2.
\end{eqnarray}
As in the SM, $\phi^\pm$ and $\sqrt{2} Im(\phi^0)$ become longitudinal 
components of $W^\pm$ and $Z$, and $\sqrt{2} Re(\phi^0) = h$ is the one 
physical Higgs boson associated with electroweak symmetry breaking.
Let $\sigma_1 = (\sigma_{1R} + i \sigma_{1I})/\sqrt{2}$ and $\sigma_2 = 
(\sigma_{2R} + i \sigma_{2I})/\sqrt{2}$, then the mass-squared matrix 
spanning $h,\sigma_{1R,2R}$ is
\begin{equation}
{\cal M}^2_R = \pmatrix {2 \lambda_0 v^2 & 2 \lambda_4 v u_1 & 
2 \lambda_5 v u_2 \cr 2 \lambda_4 v u_1 & 2 \lambda_1 u_1^2 & 
2 \lambda_3 u_1 u_2 + 2 m_{12} u_1 \cr 2 \lambda_5 v u_2 & 
2 \lambda_3 u_1 u_2 + 2 m_{12} u_1 & 2 \lambda_2 u_2^2 - m_{12} u_1^2/u_2},
\end{equation}
and that spanning $\sigma_{1I,2I}$ is
\begin{equation}
{\cal M}^2_I = \pmatrix{ -4 m_{12} u_2 & 2m_{12} u_1 \cr 2 m_{12} u_1 
& -m_{12} u_1^2/u_2}.
\end{equation}
The linear combination $(u_1 \sigma_{1I} + 2u_2 \sigma_{2I})/
\sqrt{u_1^2 + 4u_2^2}$ has zero mass and becomes the longitudinal component 
of the massive $Z_D$ gauge boson.  The orthogonal component is a 
pseudoscalar, call it $A$, with a mass given by $m_A^2 = - m_{12}(u_1^2 
+ 4u_2^2)/u_2$.  In Eq.~(14), $\sigma_{1R}$ and $\sigma_{2R}$ mix in 
general.  For simplicity, let $m_{12} = -\lambda_3 u_2$, then for 
$v^2 << u_{1,2}^2$, we obtain 
\begin{eqnarray}
&& m^2_{\sigma_{1R}} = 2 \lambda_1 u_1^2, ~~~ m^2_{\sigma_{2R}} = 2 \lambda_2 
u_2^2 + \lambda_3 u_1^2, ~~~  m^2_A = \lambda_3 (u_1^2 + 4 u_2^2),\\ 
&& m^2_h = 2 \left[ \lambda_0 - {\lambda_4^2 \over \lambda_1} - 
{2 \lambda_5^2 u_2^2 \over 2 \lambda_2 u_2^2 + \lambda_3 u_1^2} \right] v^2.
\end{eqnarray}

\noindent \underline{\it Relevance to the Diphoton Excess}~:\\
Any one of the three particles $\sigma_{1R},\sigma_{2R},A$ may be identified 
with the 750 GeV diphoton excess.  For illustration, let us consider 
$\sigma_{1R}$.  The production cross section through gluon fusion is 
given by
\begin{equation}
\hat{\sigma}(gg \to \sigma_{1R}) = {\pi^2 \over 8 m^2_{\sigma_{1R}}} 
\Gamma (\sigma_{1R} \to gg) \delta(\hat{s}-m^2_{\sigma_{1R}}).
\end{equation}
For the LHC at 13 TeV, the diphoton cross section is roughly~\cite{eeqsy15}
\begin{equation}
\sigma(gg \to \sigma_{1R} \to \gamma \gamma) \simeq (100~{\rm pb}) \times 
(\lambda_g~{\rm TeV})^2 \times B(\sigma_{1R} \to \gamma \gamma),
\end{equation}
where $\lambda_g$ is the effective coupling of $\sigma_{1R}$ to two gluons, 
normalized by
\begin{equation}
\Gamma (\sigma_{1R} \to gg) = {\lambda_g^2 \over 8 \pi} m^3_{\sigma_{1R}}, 
\end{equation}
and the corresponding $\lambda_\gamma$ comes from
\begin{equation}
\Gamma (\sigma_{1R} \to \gamma \gamma) = {\lambda_\gamma^2 \over 64 \pi} 
m^3_{\sigma_{1R}}.
\end{equation}
If $\sigma_{1R}$ decays only to two gluons and two photons, and assuming 
$\lambda_\gamma^2/8 << \lambda_g^2$, then 
\begin{equation}
\sigma(gg \to \sigma_{1R} \to \gamma \gamma) \simeq (100~{\rm pb}) \times 
(\lambda_\gamma~{\rm TeV})^2 /8,
\end{equation}
which is supposed to be about 6.2 fb from the recent 
data~\cite{atlas15,cms15}.  This means that $\lambda_\gamma \simeq 2.2 \times 
10^{-2}$ (TeV)$^{-1}$, and $\Gamma(\sigma_{1R} \to \gamma \gamma) \simeq 1$ 
MeV.

Now $\sigma_{1R}$ couples to the new scalars $\xi^{2/3},\xi^{-1/3},
\zeta^{2/3},\zeta^{-1/3},\eta^{-},\chi^{-}$ through $\sqrt{2} u_1$ 
multiplied by the individual quartic scalar couplings.  For simplicity, 
let all these couplings be the same, say $\lambda_\sigma$, and all the 
masses be the same, say $m_0$, then~\cite{hkkt16} 
\begin{equation}
\lambda_\gamma = {\alpha u_1 \lambda_\sigma \over \sqrt{2} \pi 
m^2_{\sigma_{1R}}} \left[ 6 \left( {2 \over 3} \right)^2 + 6 \left( - {1 \over 
3} \right)^2 + 2 (-1)^2 \right] f\left( {m_0^2 \over m^2_{\sigma_{1R}}} 
\right),
\end{equation}
where the function $f$ is given by
\begin{equation}
f(x) = 8 x \left[ \arctan \left( {1 \over \sqrt{4x-1}} \right) \right]^2 - 2.
\end{equation}
Let $m_0 = 700$ GeV, then $x = 0.87$ and $f=1.23$.  Hence for $u_1 = 1$ TeV 
and $\lambda_\sigma = 1.1$, the required $\lambda_\gamma \simeq 0.022$ 
(TeV)$^{-1}$ is obtained.  For this $\lambda_\sigma$, we find 
$\lambda_g = 0.128$, 
hence $\Gamma(\sigma_{1R} \to gg) \simeq 0.27$ GeV, which is below the 
energy resolution of ATLAS and CMS.  This narrow width is not favored by 
the ATLAS data, but cannot be ruled out at this time.

\noindent \underline{\it Dark Matter}~:\\
The lightest neutral particle with odd $Z_2$ is a good dark-matter 
candidate.  In this model, it could be the lightest scalar particle 
in the sector consisting of $\eta^0=(\eta_R + i \eta_I)/\sqrt{2}$ and 
$\chi^0=(\chi_R + i \chi_I)/\sqrt{2}$.  There are two sectors, the 
mass-squared matrix spanning $\eta_R,\chi_R$ is given by
\begin{equation}
{\cal M}^2_R = \pmatrix{m_\eta^2 & A \cr A & m_\chi^2 + C},
\end{equation}
and that spanning $\eta_I,\chi_I$ is
\begin{equation}
{\cal M}^2_I = \pmatrix{m_\eta^2 & B \cr B & m_\chi^2 - C},
\end{equation}
where $A,B$ come from the $\phi^0 \eta^0 (\chi^0)^*$ and $\phi^0 \eta^0 
\chi^0 (\sigma_1)^*$ couplings and $C$ from the $\chi^0 \chi^0 (\sigma_1)^*$ 
coupling.  The phenomenology of the lightest particle in this group is 
similar to that of the so-called inert Higgs doublet 
model~\cite{m06,dm78,bhr06}.  For details, see for example recent 
updates~\cite{atyy14,ikr16,dku16}.

\noindent \underline{\it Conclusion}~:\\
A new idea linking family symmetry to dark symmetry is proposed using a 
gauge $U(1)_D$ symmetry, which breaks to exactly conserved $Z_2$. The first 
and second families of quarks and leptons transform under this $U(1)_D$ so 
that their masses are forbidden at tree level.  They interact with the dark 
sector in such a way that they acquire one-loop finite masses, together 
with all three neutrinos.  The extra $Z_D$ gauge boson may have a mass 
of order a few TeV, and one particle associated with the breaking of 
$U(1)_D$ may be identified with the 750 GeV diphoton excess recently 
observed at the LHC.

\noindent \underline{\it Acknowledgement}~:~
This work was supported in part by the U.~S.~Department of Energy Grant 
No. DE-SC0008541.

\bibliographystyle{unsrt}

\end{document}